\documentclass[fleqn,12pt,twoside]{article}

\usepackage{graphics}
\usepackage{graphicx}

\usepackage{espcrc1}

\newcommand{\vs}{\,}

\newcommand{\be}{\begin{equation}}
\newcommand{\ee}{\end{equation}}
\newcommand{\ba}{\begin{eqnarray}}
\newcommand{\ea}{\end{eqnarray}}

\title{
 Power-law tailed spectra from equilibrium
}

 \author{T.~S.~Bir\'o\address[RMKI]{
  KFKI Research Institute for Particle and Nuclear Physics,
  H 1526 Budapest Pf. 49
 },
 G.~Purcsel\addressmark[RMKI],
 G\'eza Gy\"orgyi\address{Institute for
   Theoretical Physics, HAS Research Group, E\"otv\"os University,
   H-1117 Budapest, P{\'a}zm{\'a}ny s{\'e}t{\'a}ny 1/a
 },
 Antal Jakov\'ac\address{Research
       Group for Theoretical Condensed Matter of HAS and TU Budapest,
       H-1521 Budapest, Bertalan Lajos 8-10
 } and
 Zsolt Schram\address{Institute for Theoretical Physics, University of Debrecen,
         H-4010 Debrecen, Poroszlay 6/c}
 }
\begin{document}

\maketitle

%\date{{\bf \today}}
%\pacs{25.75.Nq, 05.20.Dd, 05.90.+m, 02.70.Ns}
% 25.75.Nq quark gluon plasma in heavy ion collisions
% 05.20.Dd classical kinetic theory
% 05.90.+m other topics in stat.phys. and thermodynamics 
% 02.70.Ns molecular dynamics and particle methods in computation

%\keywords{ quark matter, non-extensive thermodynamics }

\vs
\begin{abstract}
 We propose that power-law tailed hadron spectra may partially stem
 from a matter in an unconventional equilibrium state typical for
 non-extensive thermodynamics. A non-extensive Boltzmann equation,
 which is able to form such spectra as a stationary solution,
 is utilized as a rough model of quark matter hadronization.
 Basic ideas about a non-extensive simulation of the QCD equation
 of state on the lattice are presented.
\end{abstract}

%\maketitle

\section{Particle spectra}

\vs
Statistical models have been often applied to hadron physics.
Starting with Rolf Hagedorn's statistical model of meson resonances\cite{HAGEDORN},
several attempts occurred to describe hadron multiplicities in
elementary collisions by means of statistical distributions.
The very idea of a phase transition between confined and deconfined
quark matter relies on traditional equilibrium thermodynamics.
The search for quark matter began assuming a local
thermal equilibrium in an otherwise exploding fireball\cite{TOR-LECT,CSER-LECT}.
Experimental particle spectra are, however, not purely exponential:
both exponential and power-law regions have been observed in pion,
kaon and antiproton spectra.  
Although the traditional approach explains the power-law tail
at very high $p_T$ values by pQCD calculations\cite{pQCD}, the non-extensive
statistics provides a unified view for the whole spectrum.
Pion spectra from heavy ion collisions at RHIC seem to contain a power-law
part exceeding the scaled pQCD yield\cite{PHENIX-LECTURE}.
Another point can be made by inspecting the minimum bias pion $p_T$-spectrum
form RHIC AuAu collisions at 200 GeV (Fig.1 in Ref. \cite{SLIDING-SLOPE}).
A non-exponential fit can already be made at the $p_T$-region 
between $1$ and $4$ GeV. The extrapolation of this fit almost coincides
with the fit to the whole observed range between $1$ and $12$ GeV.
So one concludes that the power-law behavior is not only a very hard scale
physics. Furthermore in the non-extensive statistical approach there
is a connection between the soft properties (temperature $T$) and the
hard ones. 

\vs
The experimentally measured specific hadron spectra may reflect statistical
properties of the precursor matter\cite{STAT-MOD}. 
Fortunately transverse momentum spectra are influenced only partially, 
at their low end, by final state interactions and late resonance decay\cite{RESONANCES}. 
Since the relativistic energy is given by $E=m_T \cosh y$ with
transverse mass $m_T=\sqrt{p_T^2+m^2}$ and rapidity $y$ for a particle with mass
$m$, the best way to study statistical equilibrium distribution of
hadrons is the comparison of $m_T$-spectra at rapidity $y=0$ 
for different particles.
A universal behavior\cite{MT-SCALING} indicates that the one-particle
distributions depend on the energy only and not on all momentum
components: a basic feature of generalized and conventional
thermal distributions.

\vs

\section{Non-Extensive Boltzmann Equation}

\vs
Non-conventional distributions can be based on a non-conventional
entropy formula, which replaces the Boltzmann entropy. 
Such a formula is the Tsallis entropy, discussed
vividly in recent years.
This non-extensive thermodynamics is intended to be an effective theory
for non-equilibrium and long-range order phenomena\cite{NEXT-THERMO}.
Its canonical distribution is a power law, which occurs in particle and 
heavy-ion physics experiments.
%It is particularly interesting to produce the Tsallis distribution
%with the help of a Gamma distribution for the inverse temperature, which
%may have consequences on the e.o.s. of quark matter\cite{NEXT-MT-SCALING}. A possible
%source of such fluctuations is a multiplicative noise in the
%heat conduction\cite{HEAT-COND}. 
As nonlinear models, two generalizations of the Boltzmann equation 
have been investigated: The generalization of the product rule for
probabilities (dropping statistical independency) leads to a
non-linear Boltzmann equation\cite{NLBE}, while considering
two-particle energies composed by an extended addition rule
mounds in the non-extensive Boltzmann equation\cite{NEBE}.

\vs
The general structure of the Boltzmann equation describes the evolution
of the probability $f_1=f(\vec{p}_1)$ of a one-particle state by
considering possible transitions to and from other states:
%\be
$ \dot{f}_1 \: = \: \int \limits_{234} \, w_{1234} \,
 \left( f_{34,12} - f_{12,34} \right)$.
%\ee
Here the dot denotes a total time derivative (Vlasov operator) comprising
the essential evolution of the one-particle phase space density, $f_1$.
The indices ${1234}$ refer to two particles before and after a microcollision.
The transition probability, $w_{1234}$ 
contains conditions on conserving  momentum and energy:
\be
 w_{1234} = M^2_{1234} \, \delta((\vec{p}_1+\vec{p}_2)-(\vec{p}_3+\vec{p}_4)) \, 
                          \delta(E_{12}-E_{34}),
\ee
with $E_{12}$ total two-particle energy before and $E_{34}$ after the
collision.  The particle density factors,
$f_{12,34}$ and $f_{34,12}$ weight the transition yields for a 
$3+4\to 1+2$ and for a $1+2\to 3+4$ process, respectively.
In our approach\cite{NEBE} we keep the statistical independency, $f_{12,34}=f_1f_2$,
but generalize the energy addition formula to a nontrivial composition rule
$ E_{12} = h(E_1,E_2)$.
Rules not being a simple sum, $h(x,y)\ne x+y$, present a non-extensive
energy composition. 
Associative rules can be mapped to the simple addition\cite{MATH}:
$ X(h) = X(x) + X(y)$, unique up to a constant factor.
In each microcollision $X(E_1)+X(E_2)=X(E_3)+X(E_4)$ holds, and
the stationary solution is hence given by
$ f(p) = \frac{1}{Z} \exp(-X(E)/T)$.
A statistical dispersion relation is obtained due to the mapping of the general
composition: the total sum, $ X(E_{tot}) = \sum_i X(E_i)$ is conserved.

\begin{figure}
\begin{center}
 \includegraphics[width=0.35\textwidth,angle=-90]{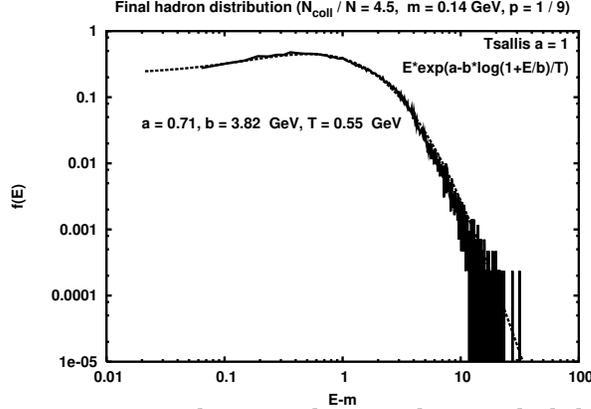}
\end{center}
\vspace{-15mm}
\caption{
  \label{NEBE-pion}
  Pion spectrum emerging by sampling with a probability $1/9$
  from the NEBE parton cascade with massless partons.
}
\end{figure}

\vs
In order to utilize such a non-extensive Boltzmann equation for
the stationary state Tsallis-distributed parton matter, we consider
an evolution from initially boosted Fermi spheres. During the
simulation we take out pairs of partons 
with a color singlet against octet probability $1/9$.
This rate turns out to be low enough not to disturb
the Tsallis distribution remarkably. Fig.\ref{NEBE-pion} shows
the resulting energy ($m_T$) distribution of such pairs.

%%%%%%%%%%%%%%%%%% --- Simulation result ---- %%%%%%%%%%%%%%%
\vs

%%%%%%%%%%%%%%%%%%%%%%%%%%%%%%%%%%%%%%%%%%%%%%%%%%%%%%%%%%%%%

\vs
It is interesting to note that an
exponentially growing mass spectrum, originally proposed by Hagedorn
and recently checked again latest experimental data in Ref.\cite{RES-SPECT},
with its famous consequence of having a limiting (or Hagedorn-) temperature
for such a system, can be reconstructed on the basis of Tsallis distributed
quark constituents. This approach\cite{ANDRE} assumes that the Tsallis distribution of
the quarks and antiquarks is folded into mesonic and baryonic distributions
of the conserved total energy satisfying $X(E)=\sum_i X(E_i)$.
The Hagedorn temperature is given by $T_H=E_c/d$.

\vs

%%%%%%%%%%%% ---- LATTICE --------- %%%%%%%%%%%%%%%%%%%%
\vs
\section{Lattice strategy}

\vs
The implementation of the Tsallis distribution in lattice field theory
can be based on the superstatistics approach\cite{SUPER-STAT}.
The Tsallis expectation value of an observable $\hat{A}[U]$ over  lattice field
configurations $U$ is of interest. It may include
the timelike link length, say  on the power $v$:
$\hat{A}=\theta^{\:v}A$.
The Tsallis expectation value then is an average over all possible $a_t$
link lengths according to a Gamma distribution of $a_t/a_s$.
We obtain:
\be
 \langle A \rangle_{TS} \, =  \, \frac{1}{Z_{TS}} \frac{c^c}{\Gamma(c)}
 \int\!d\theta\: \theta^{\: c-1} e^{-c\:\theta} \int {\cal D}U A\left[U\right]
 \theta^{\:v} e^{-S\left[\theta,U\right]}
 \label{TS-EXP}
\ee
with $Z_{TS}$ obtained by requiring $\langle 1 \rangle_{TS}=1 $.
%\be
% Z_{TS} \, =  \, \frac{c^c}{\Gamma(c)}
% \int\!d\theta\: \theta^{\: c-1} e^{-c\:\theta} \int {\cal D}U
% e^{-S\left[\theta,U\right]}.
%\ee
The $\theta$ dependence of the lattice gauge action is known long.
Due to the time derivatives 
the electric(''kinetic'') part scales like $a_ta_s^3/(a_t^2a_s^2)=a_s/a_t$,
and the magnetic (''potential'') part like $a_ta_s^3/(a_s^2a_s^2)=a_t/a_s$.
%\footnote{This generalizes to all lattice field actions: kinetic and mass
%terms scale like $1/\theta$, potential terms like $\theta$.}.
This leads to the following expression for the general lattice action:
%\be
$ S\left[\theta,U\right] = a \: \theta + b / \theta$,
% \label{SLAT}
%\ee
where 
$a=S_{ss}[U]$ sums space-space, and $b=S_{ts}[U]$ time-space oriented plaquettes. 
In the $c \rightarrow \infty$ limit the scaled Gamma distribution approximates
$\delta(\theta-1)$, (its width narrows extremely, while its integral
is normalized to one), and one gets back the traditional lattice
action $S=a+b$, and the traditional averages.
For finite $c$, one can exchange
the $\theta$ integration and the configuration sum (path integral) and
obtains exactly the power-law-weighted expression: 
%\be
$ \langle A \rangle_{TS} \, = \, 
    {\int{\cal D}U \: W_{v, c}\left[U\right]} A\left[U\right] \: 
    / \: {\int{\cal D}U \: W_{0, c}\left[U\right]},
$
%\ee
with the Gamma fluctuating time-link averaged general weight factor,
\be
  W_{v,c} = \frac{c^c}{\Gamma(c)} \int d\theta \: \theta^{\:v+c-1}
  e^{-c\: \theta} e^{-S\left[\theta,U\right]}.
\ee
The $\theta$ integration can be carried out analytically using 
the replacement $\theta = e^t\,\sqrt{b/(a+c)}$. 
The result contains the $K$ Bessel function:
\be
  W_{v,c} = \frac{c^c}{\Gamma(c)} 
  \left(\frac{b}{a+c}\right)^{\frac{c+v}{2}}
  2 \: K_{v+c}\left(\, 2\sqrt{b(a+c)} \, \right).
  \label{TS-WEIGHT}
\ee
The K-Bessel function has an exponentially decreasing asymptotics, so
we are in principle able to utilize known Monte Carlo techniques 
in order to calculate Tsallis expectation values. On the other hand we
cannot simply use old data, produced according to the weight $e^{-(a+b)}$,
because the argument of the K-Bessel function is not $a+b$. This makes
it necessary to redo lattice calculations --  but only with  a slightly
increased effort.

\vs
{\bf Acknowledgment}
\vs

Enlightening discussions with H. Markum, J. Pol\'onyi
and J. Zim\'anyi are gratefully acknowledged. 
This work was supported by the Hungarian National Research Fund
OTKA (T049466, T046925).

%%%%%%%%%%%%%%%%%%%% REFERENCES %%%%%%%%%%%%%%%%%%

%%%%%%%%%%%%%%%%%%%%%%%%%%%%%%%%%%%%%%%%%%%%%%%%%%%%%%%%%%%%%%%%%%%%%%%%%%%%%%%%%%%

\end{document}